\documentclass[final,leqno]{siamltexmm}
\usepackage{amsmath,amssymb}

\usepackage{graphicx}
\usepackage{dcolumn}
\usepackage{bm}
\usepackage{subeqnar}
\usepackage{overpic,color,rotating}
\usepackage{overpic}
\usepackage{gensymb}

\graphicspath{{figures/}}

\title{Dynamic Mode Decomposition for Financial Trading Strategies}

\author{Jordan Mann\thanks{Harvard University, Cambridge, MA.} 
\and J. Nathan Kutz$^\ddagger$\thanks{Department of Applied Mathematics, University of Washington, Seattle, WA. 98195-2420.  $^\ddagger$ (email: {kutz@uw.edu})  Questions, comments, or corrections to this document may be directed to that email address.}}

\begin{document}
\maketitle
\newcommand{\slugmaster}{%
\slugger{sifin}{}{}{}{}}

\begin{abstract}
We demonstrate the application of an algorithmic trading strategy based upon the recently developed dynamic mode decomposition (DMD) on
portfolios of financial data.  The method is capable of  characterizing complex dynamical systems, in this case financial market dynamics, in an equation-free manner by decomposing the state of the system into low-rank terms whose temporal coefficients in time are known.
By extracting key temporal coherent structures (portfolios) in its sampling window, it provides
a regression to a best fit linear dynamical system, allowing for a predictive assessment of the market dynamics
and informing an investment strategy.  The data-driven analytics capitalizes on stock market patterns, either real or perceived, to inform buy/sell/hold investment decisions.  Critical to the method is an associated learning algorithm that optimizes the sampling and prediction windows of
the algorithm by discovering trading hot-spots. The underlying mathematical structure of the algorithms is rooted in methods from nonlinear dynamical systems and 
shows that the decomposition is an effective mathematical tool for data-driven discovery of market patterns.
\end{abstract}



\section{Introduction}

Algorithmic trading (alg trading) schemes are of growing importance in modern 
day financial investment strategies.  In 2006, for instance,
it was estimated that one third of all European Union and United States stock, along with 40\% of the
London Stock Exchange, were executed by trading algorithms.  The advent of high-frequency trading, which
was estimated in 2009 to accounted for 60-73\% of all US equity trading volume~\cite{alg1,alg2},
has only added to the confluence of automated traders.  Alg trading is driven by mathematical models and modern data-driven analytics which seek to capitalize on stock market patterns, either real or perceived, to inform buy/sell/hold investment decisions.  The underlying mathematical structure of the algorithms is typically rooted in sophisticated statistical and probabilistic computational tools, thus providing risk measures and confidence intervals in the decision making process.  
We develop a trading scheme based upon ideas from nonlinear dynamical systems that capitalizes on
evanescent signals in financial market data.   Specifically, we apply the {\em dynamic mode 
decomposition} (DMD)~\cite{DataBook,DMD0,DMD1,DMD4,DMD2,DMD3,DMD5}, which is an emerging data analysis tool capable of integrating the power
of time-series analysis with Principal Component Analysis (PCA),  to financial data and portfolios of holdings.  The
DMD method extracts key temporal coherent structures (portfolios) in its sampling window and provides
a regression to a best fit linear dynamical system, allowing for a predictive assessment of the current market
and informing an investment strategy.  

The viewpoint advocated here assumes the stock market to be a complex, dynamical
system that exhibits non-stationary, multi-scale phenomenon.    
But unlike standard dynamical systems methods, 
the DMD method does not enforce or prescribe an underlying dynamical model.  Rather,
it is an {\em equation-free} method whereby the dynamics are reconstructed directly from the data sampled
over a specified window of time.  Specifically, the DMD decomposes stock portfolio data into low-rank
features that behave with a prescribed time dynamics.   In the process, the least-square fit 
linear dynamical system allows one to predict short-time future states of the system. 
These short-time predictions can be used to formulate successful trading strategies by identifying 
transient, evanescent signals in the financial data.  The method is adaptive, updating its
decomposition as the market changes in time.  It also uses a learning (machine learning) 
algorithm to track optimal sampling and prediction windows as these change much more slowly in time
and in different market sectors.
The method is applied to several markets (e.g. technology, bio tech, transport, banks) and is demonstrated
to produce a robust trading strategy.

In a broader context, modeling of multi-scale systems, both in time and space, pervade modern theoretical and computational efforts across the engineering, biological and physical sciences.  Driving innovations are methods and algorithms that circumvent the significant challenges in efficiently connecting micro-scale to macro-scale effects that are separated potentially by orders of magnitude spatially and/or temporally.
As such, the origins of the DMD method, which arose from pioneering work connecting the Koopman operator to 
dynamical systems theory~\cite{Mezic2004,Mezic2005}, are associated with the fluid dynamics community and the modeling of complex flows~\cite{DMD1,DMD4}.  Its growing success stems from the fact that it is an {\em equation-free}, data-driven method~\cite{DataBook} capable of providing accurate assessments of the spatio-temporal coherent structures in a complex system, or short-time future estimates, thus potentially allowing for control protocols to be enacted simply from data sampling.  In the context of financial data, the analogy of a spatial structure would be a portfolio of
stock holdings.

The DMD method exhibits many features of ARIMA (autoregressive integrated moving averages) models and
key extensions like SARIMA (Seasonal ARIMA)~\cite{arima}.  However, the DMD method by construction correlates both
temporal and spatial data simultaneously and extracts low-rank features that a time-series or PCA analysis
cannot individually.  The DMD algorithm also allows one to adapt the frequency and duration (sampling window) of the market data collection
to sift out information at different time scales, making different trading strategies 
(e.g. high-frequency, daily trading, long-term trading etc) possible.  Indeed, one can use an iterative refinement process 
to optimize the snapshot sampling window for predicting the future market.
A critical innovation of DMD is its ability to handle transient phenomenon and non-stationary data, which
are typically weaknesses  of SVD-based techniques.  One can also build upon recent innovations
in multi-resolution DMD for mining for data features at different timescales~\cite{mrdmd}.

The paper is outlined as follows:  In Sec.~\ref{sec:dmd} the basic DMD theory is outlined with an emphasis on its dynamical approximation of data.  This is followed in Sec.~\ref{sec:dmd_alg} by the development of the DMD algorithm used in the subsequent applications.  The application of the DMD to trading strategies 
is outlined in Sec.~\ref{sec:dmd_trade} with various subsections demonstrating the
efficacy of the method.  The paper is concluded in Sec.~\ref{sec:conclusion} with an outlook of 
the method as a modern data tool for financial analysis.

\section{Dynamic Mode Decomposition:  An Equation-Free Architecture}
\label{sec:dmd}

The origins of the DMD method are associated with the fluid dynamics community and the modeling of complex flows.
As with financial data, complex fluids exhibit dynamics that are difficult for model-based approaches
to accurately predict.  Unlike fluids however, financial markets have no known
{\em laws of nature}, thus necessitating statistical modeling approaches.  DMD is a natural tool for
financial modeling as its growing success stems from the fact that it is an {\em equation-free}, data-driven method capable of providing accurate assessments of spatial-temporal coherent structures in a given complex system, or short-time future estimates of such a systems, thus allowing for state reconstruction and control protocols to be enacted simply from sampling.  

To be more precise, one may consider the DMD as an equation-free way to approximate the nonlinear dynamics of a dynamical or complex system.  We can formulate mathematical framework of 
DMD by considering the governing set of differential equations:
\begin{equation}
  \frac{d{\bf x}}{dt} = {N}({\bf x},t;{\mathbf{\mu}}) \, ,
  \label{eq:U}
\end{equation}
where ${\bf x}$, in the interpretation here, is a portfolio of companies that are selected
for evaluation.  The function $N(\cdot)$ is an unknown, dynamical process that
is generally nonlinear and time-dependent.  Further, it may depend on a set of
bifurcation parameters, ${\mathbf{\mu}}$, that can alter the underlying dynamics.

In addition to the governing equations, both measurements of the system, denoted by $G(\cdot)$,
\begin{equation}
  G({\bf x},t_k)=0 \, ,
\end{equation}
where $k=1,2,\cdots,M$ for a total of $M$ measurement times,
and initial conditions are prescribed
\begin{equation}
 {\bf x}(0)={\bf x}_0.
\end{equation} 
In applications of DMD to engineering and physical sciences,  typically ${\bf x}$ is an $n$-dimensional vector ($n\gg 1$) that arises from either 
discretization of a complex system, or in the case of applications such as video streams, it is the total number of
pixels in a given frame.  The governing equations and initial condition specify a well-posed
initial value problem.  The inclusion of measurements $G({\bf x},t_k)$, let's say $M$ of them,
make the system overdetermined.  By including model error along with noisy measurements,
one can formulate an optimal predictive strategy using the data-assimilation 
framework and Kalman filtering innovations~\cite{DataBook}.

Since in general the solution of governing nonlinear evolution (\ref{eq:U}) is not possible to construct,
numerical solutions are used to evolve to future states.  In the DMD framework, recall that the equation-free viewpoint  assumes that the right-hand side governing the dynamics, ${N}({\bf x},t;{\mathbf{\mu}})$, is unknown.  Thus the snapshot measurements and initial conditions alone are used to approximate the dynamics and predict the future state.    The DMD procedure thus constructs the proxy, approximate linear evolution
\begin{equation}
  \frac{d \tilde{\bf x}}{dt} = {\bf A} \tilde{\bf x}
\end{equation}
with $\tilde{\bf x}(0) = \tilde{\bf x}_0$ and whose well-known solution~\cite{boyce} is
\begin{equation}
   \tilde{\bf x}(t)=  \sum_{k=1}^K b_k \psi_k \exp(\omega_k t) \, 
   \label{eq:omegaj}
\end{equation}
where $\psi_k$ and $\omega_k$ are the eigenfunctions  and eigenvalues 
of the matrix ${\bf A}$.  Of particular importance for finance is the interpretation of
this solution.  In particular, portfolio (DMD) modes with a positive real
part of $\omega_k$ are exponentially growing solutions, thus making money, while those with negative
real part are exponentially decreasing and loosing money.  Investment strategies
will be based upon the values of $\omega_k$ achieved in the DMD decomposition.

The ultimate goal in the DMD algorithm is to optimally construct the matrix ${\bf A}$ so
that the true and approximate solution remain optimally close in a least-square 
sense:  
\begin{equation}
\| {\bf x}(t) - \tilde{\bf x}(t) \| \ll 1. 
\end{equation}
Of course, the optimality of the approximation holds only over the sampling window where ${\bf A}$ is constructed, but the approximate solution can be used to not only make future state predictions, but also decompose the dynamics 
into various time-scales since the $\omega_k$ are prescribed.  Moreover, the DMD makes use
of low-rank structure so that the total number of modes, $K\ll N$, allows for dimensionality reduction of the complex system.

\section{The DMD Decomposition and Algorithm}
\label{sec:dmd_alg}

The DMD method provides a decomposition of data into
a set of dynamic modes that are derived from snapshots or measurements of a given system in time.   The mathematics underlying the extraction of dynamic information from time-resolved snapshots  is closely related to the idea of the Arnoldi algorithm~\cite{DMD1}, one of the workhorses of fast computational solvers.   The data collection process involves two parameters:
\begin{subeqnarray}
  &&  N = \mbox{number of companies in a given portfolio} \nonumber \\
  &&  M= \mbox{number of data snapshots taken} \nonumber
\end{subeqnarray}
Originally the algorithm was designed to collect data at regularly spaced intervals of time, e.g. the daily opening
price of a stock.  However, new innovations allow for both sparse market~\cite{cdmd} and temporal~\cite{Tu2014ef} collection of data as well as irregularly spaced collection times~\cite{DMD5}.  To illustrate the algorithm, we consider regularly spaced sampling in time:
\begin{equation}
  \mbox{data collection times}: \,\,\,\, t_{m+1} = t_{m} + \Delta t 
\end{equation}
where the collection time starts at $t_1$ and ends at $t_M$, and the
interval between data collection times is $\Delta t$.  In optimizing the method, the total number of snapshots is
varied to determine best performance.

The data snapshots are arranged into an $N\times M$ matrix
\begin{equation}
  {\bf X} = \left[ {\bf x}(t_1) \,\,\, {\bf x}(t_2) \,\,\, {\bf x}(t_3) \,\,\, \cdots \,\,\,  {\bf x}(t_M)    \right]
\end{equation}
where the vector ${\bf x}$ are the $N$ measurements of the state variable of the system of interest at the data collection points.  Specifically, each component of the vector ${\bf x}$ is a company that comprises the portfolio of companies to be evaluated.   The objective is to mine the data matrix ${\bf X}$ for important dynamical information.  For the purposes of the DMD method, the following matrix is also defined:
\begin{equation}
  {\bf X}_j^{k} = \left[  {\bf x}(t_j) \,\,\, {\bf x}(t_{j+1}) \,\,\,  \cdots \,\,\, {\bf x}(t_k)    \right]
\end{equation}
Thus this matrix includes columns $j$ through $k$ of the original data matrix.

The DMD method approximates the modes of the so-called {\em Koopman operator}.
The Koopman operator is a linear, infinite-dimensional operator that represents
nonlinear, infinite-dimensional dynamics without linearization~\cite{DMD4,mezic2},
and is the adjoint of the Perron-Frobenius operator\index{Perron-Frobenius operator}.  
The method can be viewed as computing, from the experimental data, the eigenvalues and eigenvectors (low-dimensional
modes) of a linear model that approximates the underlying dynamics, even if the dynamics
is nonlinear.  Since the model is assumed to be linear, the decomposition gives
the growth rates and frequencies associated with each mode.  If the underlying model
is linear, then the DMD method recovers the leading eigenvalues and eigenvectors 
normally computed using standard solution methods for linear differential equations.

Mathematically, the Koopman operator ${\bf A}$ is a linear, time-independent operator ${\bf A}$
such that\index{Koopman operator}
\begin{equation}
   {\bf x}_{j+1} = {\bf A} {\bf x}_j
   \label{eq:Koopman}
\end{equation}
where $j$ indicates the specific data collection time and ${\bf A}$ is the linear operator
that maps the data from time $t_{j}$ to $t_{j+1}$.  The vector ${\bf x}_j$ is an $N$-dimensional
vector of the data points collected at time $j$.  The computation of the Koopman operator
is at the heart of the DMD methodology.  As already stated, the mapping over $\Delta$ is
linear even though the underlying dynamics that generated ${\bf x}_j$ may be nonlinear.  It should
be noted that this is different than linearizing the dynamics.

To construct the appropriate Koopman operator that best represents the data collected,
the matrix ${\bf X}_1^{M-1}$ is considered:
\begin{equation}
  {\bf X}_{1}^{M-1} = \left[ {\bf x}_1 \,\,\, {\bf x}_2 \,\,\, {\bf x}_3 \,\,\, \cdots \,\,\, {\bf x}_{M-1} \right] \, .
\end{equation}
Making use of (\ref{eq:Koopman}), this matrix reduces to
\begin{equation}
  {\bf X}_{1}^{M-1} = \left[ {\bf x}_1 \,\,\, {\bf A}{\bf x}_1 \,\,\, {\bf A}^2{\bf x}_1 \,\,\, \cdots \,\,\, 
    {\bf A}^{M-2}{\bf x}_{1} \right] \, .
    \label{eq:kry1}
\end{equation}
Here is where the DMD method connects to Krylov subspaces\index{Krylov subspace} and the Arnoldi algorithm\index{Arnoldi algorithm}.
Specifically, the columns of ${\bf X}_{1}^{M-1}$ are each elements in a Krylov space.
This matrix attempts to fit the first $M-1$ data collection points using the Koopman operator (matrix)
${\bf A}$.  In the DMD technique, the final data point ${\bf x}_M$ is represented, as best as possible,
in terms of this Krylov basis, thus
\begin{equation}
  {\bf x}_M = \sum_{m=1}^{M-1} b_m {\bf x}_m + {\bf r}
  \label{eq:kry2}
\end{equation}
where the $b_m$ are the coefficients of the Krylov space vectors and ${\bf r}$ is the residual
(or error) that lies outside (orthogonal to) the Krylov space.  Ultimately, this best fit to the data
using this DMD procedure will be done in an $L^2$ sense using a pseudo-inverse, i.e. the residual
${\bf r}$ is minimized in the DMD procedure.

Before proceeding further, it is at this point that the data matrix $ {\bf X}_{1}^{M-1}$
in (\ref{eq:kry1}) should be considered further.  In particular, our dimensionality
reduction methods look to take advantage of any low-dimensional structures in
the data.    To exploit this, the SVD of (\ref{eq:kry1}) is computed~\cite{tre}:
\begin{equation}
  {\bf X}_{1}^{M-1} = {\bf U} {\bf \Sigma} {\bf V}^*
  \label{eq:kry3}
\end{equation}
where $*$ denotes the conjugate transpose, 
${\bf U}\in {\mathbb{C}}^{N\times K}$, ${\bf \Sigma}\in {\mathbb{C}}^{K\times K}$
and ${\bf V}\in {\mathbb{C}}^{M-1\times K}$.  Here $K$ is the reduced SVD's
approximation to the rank of ${\bf X}_{1}^{M-1}$.  If the data matrix is full rank
and the data has no suitable low-dimensional structure, then the DMD method
fails immediately.  However, if the data matrix can be approximated by
a low-rank matrix, then DMD can take advantage of this low dimensional structure
to project a future state of the system.  Thus once again, the SVD plays the
critical role in the methodology.

Armed with the reduction (\ref{eq:kry3}) to (\ref{eq:kry1}), we can return
to the results of the Koopman operator and Krylov basis (\ref{eq:kry2}).  In particular,
we will consider constructing the matrix ${\bf A}$ that gives the best approximation
\begin{equation}
  {\bf A} {\bf X}_{1}^{M-1} \approx {\bf X}_{2}^{M} \, .
\label{eq:A1A2}
\end{equation}
But by using (\ref{eq:kry2}), the right hand side of this equation can be written
in the form
\begin{equation}
  {\bf X}_{2}^{M}  = {\bf X}_{1}^{M-1} {\bf S} + {\bf r} e_{M-1}^*
\end{equation}
where $e_{M-1}$ is the $(M-1)$th unit vector and
\begin{equation}
  {\bf S} = \left[     
  \begin{array}{ccccc}
     0 & \cdots & & 0 & b_1 \\
     1 & \ddots & & 0 & b_2 \\ 
     0 & \ddots & \ddots &   &  \vdots \\
       &  \ddots & \ddots &  0 & b_{M-2}  \\
      0 &    \cdots        &   0  & 1 & b_{M-1}
     \end{array}
  \right].
\end{equation}
Recall that the $b_j$ are the unknown coefficients in (\ref{eq:kry2}). 

A key observation is that some of the eigenvalues of ${\bf A}$ can 
be determined by a similarity transformation to the matrix $\tilde{\bf A}={\bf U} {\bf A} {\bf U}^*$.  Thus we 
approximate the unknown Koopman operator
${\bf A}$ with $\tilde{\bf A}$, making the DMD method similar to the Arnoldi algorithm and
its approximations to the Ritz eigenvalues~\cite{DMD1}.  Using equation (\ref{eq:A1A2}) with (\ref{eq:kry3}) gives
\begin{equation}
  \tilde{\bf A} = {\bf U}^*  {\bf X}_{2}^{M} {\bf V} {\bf \Sigma}^{-1} \, .
\end{equation}
Recall that the matrices ${\bf U}$, ${\bf \Sigma}$ and ${\bf V}$ arise from
the SVD reduction of ${\bf X}_{1}^{M-1}$ in (\ref{eq:kry3}).  This is done
in practice since in the fluids literature where DMD was developed, the matrix ${\bf A}$
is extremely high-dimensional and computing it directly is computationally challenging.
The matrix $\tilde{\bf A}$, however, is of reduced dimension and can be
computed relatively easily.

Consider then the eigenvalue problem associated with $ \tilde{\bf A} $:
\begin{equation}
    \tilde{\bf A} {\bf y}_k = \mu_k {\bf y}_k  \hspace*{0.5in} k=1, 2, \cdots, K
\end{equation}
where $K$ is the rank of the approximation we are choosing to make.
The eigenvalues $\mu_k$ capture the time dynamics of the discrete Koopman map ${\bf A}$ 
as a $\Delta t$ step is taken forward in time.  
These eigenvalues and eigenvectors can be related back 
to the similarity transformed original eigenvalues and eigenvectors of ${\bf S}$
in order to construct the DMD modes:
\begin{equation}
  \psi_k = {\bf U} {\bf y}_k \, .
\end{equation}
With the low-rank approximations of both the eigenvalues and eigenvectors in
hand, the projected future solution can be constructed for all time in the future.
By first rewriting for convenience $\omega_k=\ln(\mu_k)/\Delta t$ (recall that
the Koopman operator time dynamics is linear), then the approximate solution
at all future times, $ {\bf x}_{\mbox{\tiny DMD}}(t)$, is given by
\begin{equation}
  {\bf x}_{\mbox{\tiny DMD}}(t) = \sum_{k=1}^{K} b_k(0) \psi_k ({\bf x}) \exp(\omega_k t) = {\bf \Psi} 
  \mbox{diag} (\exp(\omega t) {\bf b} 
  \label{eq:dmd_sol}
\end{equation}
where $b_k(0)$ is the initial amplitude of each mode, ${\bf \Psi}$ is the matrix whose
columns are the eigenvectors $\psi_k$, $\mbox{diag}(\omega t)$ is a diagonal
matrix whose entries are the eigenvalues $\exp(\omega_k t)$, and ${\bf b}$ is a
vector of the coefficients $b_k$.\index{eigenvalues}\index{eigenfunctions}\index{POD}

It only remains to compute the initial coefficient values $b_k(0)$.  If we consider the
initial snapshot (${\bf x}_1$) at time zero, let's say, then (\ref{eq:dmd_sol}) gives
${\bf x}_1 ={\bf \Psi} {\bf b}$.  This generically is not a square matrix so that
its solution 
\begin{equation}
  {\bf b} = {\bf \Psi}^+ {\bf x}_1
\end{equation}
can be found using a pseudo-inverse.  Indeed, ${\bf \Psi}^+$ denotes the Moore-Penrose
pseudo-inverse\index{pseudo-inverse} that can be accessed in MATLAB via the {\bf pinv} command.  As already
discussed in the compressive sensing section, the pseudo-inverse is equivalent to
finding the best solution ${\bf b}$ the in the least-squares (best fit) sense.  This is equivalent
to how DMD modes were derived originally.

Overall then, the DMD algorithm presented here takes advantage of low dimensionality
in the data in order to make a low-rank approximation of the linear mapping that
best approximates the nonlinear dynamics of the data collected for the system.  Once
this is done, a prediction of the future state of the system is achieved for all time.  Unlike
the POD-Galerkin method, which requires solving a low-rank set of dynamical quantities to
predict the future state, no additional work is required for the future state prediction outside
of plugging in the desired future time into (\ref{eq:dmd_sol}).  Thus the advantages
of DMD revolve around the fact that (i) no equations are needed, and (ii) the future state
is known for all time (of course, provided the DMD approximation holds).

The algorithm is as follows:\\

\noindent
(i) Sample data at $N$ prescribed locations $M$ times.  The data snapshots
should be evenly spaced in time by a fixed $\Delta t$. This gives the data
matrix ${\bf X}$.\\

\noindent
(ii) From the data matrix ${\bf X}$, construct the two sub-matrices ${\bf X}_{1}^{M-1}$  and
${\bf X}_{2}^{M}$.\\

\noindent
(iii) Compute the SVD decomposition of ${\bf X}_{1}^{M-1}$. \\

\noindent
(iv) The matrix $\tilde{\bf S}$ can then be computed and its eigenvalues
and eigenvectors found.\\

\noindent
(v) Project the initial state of the system onto the DMD modes using
the pseudo-inverse.\\

\noindent
(vi) Compute the solution at any future time using the DMD modes along
with their projection to the initial conditions and the time dynamics computed
using the eigenvalue of $\tilde{\bf A}$.\\

Recall that the future projection of step (vi), particularly the growth or demise of
a portfolio, is largely based upon the real part of the eigenvalues $\omega_k$
computed in step (iv).  One interpretation of DMD is that we ultimately invest in
eigenvalue distributions and their associated DMD modes.  The implementation
of this algorithm in MATLAB can be found in Ref.~\cite{DataBook}.

\section{Financial Trading with DMD}
\label{sec:dmd_trade}

With the DMD theory in hand, we can turn our attention to building a trading algorithm
which capitalizes on the predictions (\ref{eq:dmd_sol}) of the theory.   The trading
algorithm is parametrized by two key (integer) parameters:
\begin{subeqnarray}
  &&  m = \mbox{number of past days of market snapshot data taken} \nonumber \\ 
  &&  \ell = \mbox{number of days in the future predicted} \nonumber .
\end{subeqnarray}
Specifically, we will refer to the DMD prediction (\ref{eq:dmd_sol}) with the notation
\begin{equation}
  {\bf x}_{\mbox{\tiny DMD}} (m,\ell)
  \label{eq:xml}
\end{equation}
to indicate the past $m$ number of days that are used to predict $\ell$ days in the future.  This
allows us to specify both the market sampling window and how far in the future we are predicting.
Our objective is to use historical data to determine suitable combinations of $(m,\ell)$ that
give the best predictive value.  In particular, we look for what we term {\em trading hot-spots}, or
regions of $(m,\ell)$ where the predictions are quite good.  In the following subsection, we focus on
two key steps:  (i) a training step in the algorithm for determining $(m,\ell)$ and (ii) implementation
of trading based upon the results.

\subsection{Trading Algorithm and Training}

The training algorithm allows us to learn more about which inputs work best with DMD analysis for given sectors. The training alg looks over a historic time period, whether that is 100 days or 10 years, in order to determine
the best choices of $(m,\ell)$.   We consider all possible combinations of $(m,\ell)$ and their associated
success (prediction) rates.   Since we are using historical data, we can compare the DMD prediction
with known market activity.   Specifically, we evaluate if the DMD predicts that the market increases or decreases, 
and we compare that to actual market activity.   In what follows, we focused our attention on daily trading since that data was readily available through Yahoo!~\cite{yahoo}.  We set limits that were suitable for the DMD algorithm, letting $m=1, 2, \cdots, 25$ days and allowing $\ell=1, 2, \cdots 10$ days.  As will be shown in the results, these appear to be 
reasonable and effective values for determining the best combinations of $(m,\ell)$.

When we looked at the training alg over the last 10 years we could see consistent trading hot-spots across sectors. Most hot-spots would look at the last 8-10 days of prices to make the best prediction of the price in 4-5 days time. Hot-spots that had success rates greater than 50\% were promising because they would likely make money over time.   The
information gathered about hot-spots allowed us to create a trading algorithm that would enter stock market positions using results from DMD analysis each day and the solution (\ref{eq:xml}).

There were a few trading algorithms that we used throughout the project.   However all of them used the same three basic assumptions:  (i) initial capital of \$1 million, (ii) transaction costs of \$8 for each position, and (iii) all money would be invested evenly across all companies in the portfolio.   We had the flexibility to use any company, providing they had been publicly trading for the time frame we were using, and we were able to combine as many companies as we wished. For illustrative purposes, we tended to use ten companies as a proxy, or representation, of each sector considered.
The initial daily trading alg took given inputs $(m,\ell)$ for the sampling window and prediction window, and ran the DMD analysis each day.   Specifically, trading was performed using the DMD hot-spot prediction windows and capital was divided equally among all companies in the portfolio.  After we entered the position for a given duration, we calculated how much money would have been made or lost by using historical data.  We also re-invested gains over the duration of the trade period.  After the alg had completed we compared our results to buying the benchmark, S\&P 500, and also compared it to buying and holding the individual stocks.  
Note that effects of slippage, for instance, have been ignored in the trading scheme.  However, the \$8 per trade is a conservative (high) estimate of trading costs that should offset such effects.

The second trading alg we created did not use any information that one wouldn't have if they wanted to trade today; hence it was as realistic as possible. The algorithm would start trading at day 101 because it would continuously use the previous 100 days to find the optimal trading inputs from the training alg. Therefore the training alg would be used on a sliding 100 day window prior to the day the trading alg was executed.  It would update its results daily using the previous 100 days.  Throughout our research we found that most sectors had obvious, even prominent, hot-spots.
However some sectors didnÕt have any clear hot-spot, and they tended to be the sectors that underperformed in the trading alg. 

With this in mind, we created a third trading alg that looked into whether the inputs with the maximum success rate were within a larger hot-spot region or isolated instances and likely to have performed well over the last 100 days due to randomness. To do this, the trading alg found out what inputs had the maximum success rate over the last 100 days, and then it looked at the surrounding inputs to see if the mean success rate of all 9 neighboring $(m,\ell)$ were above a threshold.   If a hot-spot region was found, then it would be classified as a hot-spot and a trade would be executed.  Otherwise the trading alg would hold the money until a hot-spot appeared at a later date. When implementing this strategy, we used a hotspot threshold of 53\% so that we could be confident there truly was a hot-spot. 
This is perhaps the most robust of the trading strategies as we demonstrated that markets with hot-spot regions where
quite amenable to the DMD strategy.  Indeed, all market sectors showing a strong hot-spot generated
significant capital gains with the trading alg. 

%

\begin{figure}[t]
\begin{center}
\begin{overpic}[width=0.95\textwidth]{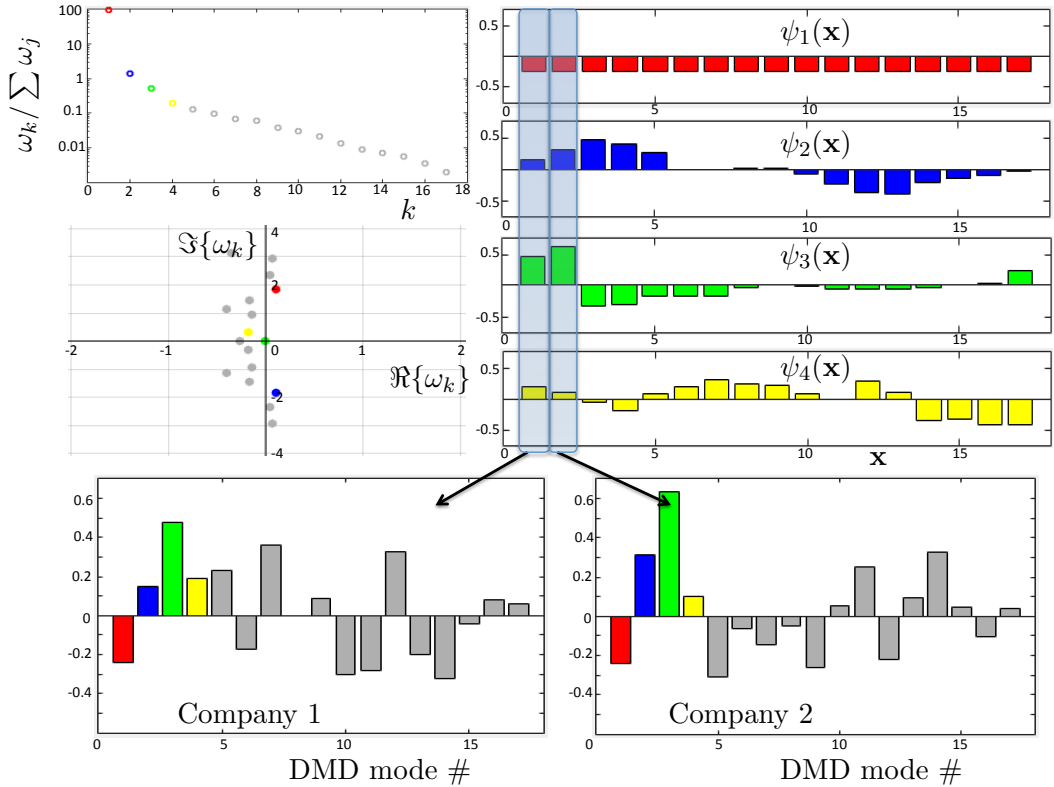}
\normalsize
\put(70,69){$\psi_1 ({\bf x})$}
\put(70,59){$\psi_2 ({\bf x})$}
\put(70,49){$\psi_3 ({\bf x})$}
\put(70,39){$\psi_4 ({\bf x})$}
\put(1,59){\rotatebox{90}{$\omega_k/\sum \omega_j$}}
\put(16,50){$\Im \{ \omega_k \}$}
\put(35,38){$\Re \{ \omega_k \}$}
\put(36,53){$k$}
\put(16,8){Company 1}
\put(60,8){Company 2}
\put(26,3){DMD mode \#}
\put(70,3){DMD mode \#}
\put(78,31){${\bf x}$}
\end{overpic}
\end{center}
\caption{\label{fig1}  DMD decomposition of an 18-day sampling of portfolio data in the biotech and healthcare sector (17 companies):
'DHI' 'LEN' 'PHM' 'TOL' 'NVR' 'HD' 'LOW' 'SHW' 'ONCS' 'BIIB' 'AMGN' 'CELG' 'GILD' 'REGN' 'VRTX' 'ALXN' 'ILMN'.
The top left panel shows, on a log scale, the percentage of information captured in each mode from the SVD
decomposition (\ref{eq:kry3}) ($\sigma_j/\sum \sigma_k$ where $\sigma_k$ are
the diagonal elements of $\Sigma$).  The data, which is color coded throughout the figure, 
is shown to be dominated by a few leading modes (colored red,
blue, green and yellow in order of variance contained). The middle left shows the 17 eigenvalues $\omega_k$ of each mode used in the solution (\ref{eq:dmd_sol}).  Eigenvalues with $\Re\{\omega_i\}>0$ represent
growth modes.   The 4 top right panels shows the leading DMD modes ($\psi_k ({\bf x})$) and their composition from the 17 companies selected.  The first mode (red) shows the ``background" portfolio,
or average price of stocks, over the sampling window.  The decomposition of the first and second companies
on the 17 DMD modes is shown in the bottom two panels where the first four modes are highlighted.}
\end{figure}

\begin{figure}[t]
\begin{center}
\begin{overpic}[width=0.9\textwidth]{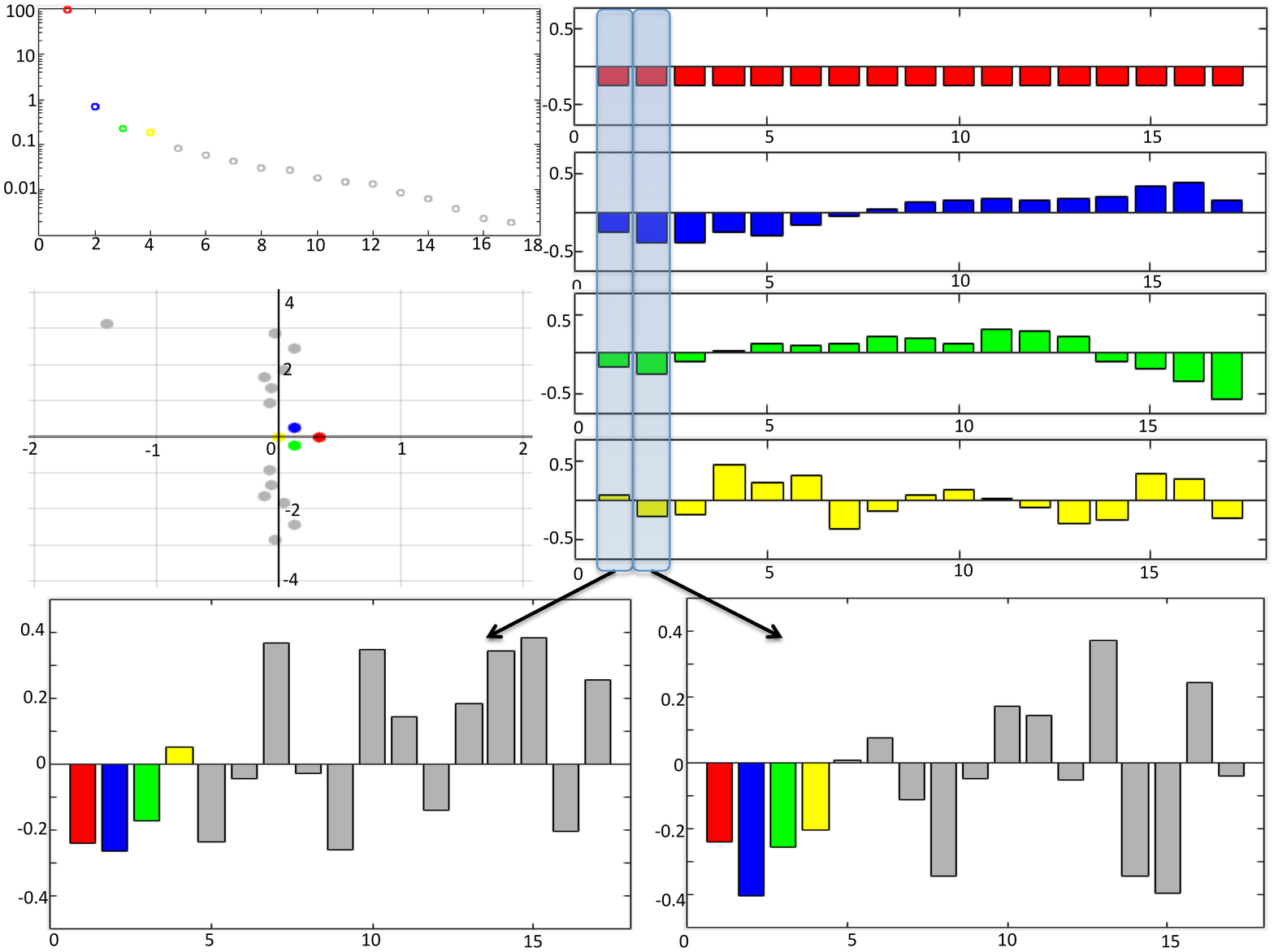}
\normalsize
\put(63,69){$\psi_1 ({\bf x})$}
\put(63,59){$\psi_2 ({\bf x})$}
\put(63,49.2){$\psi_3 ({\bf x})$}
\put(63,39){$\psi_4 ({\bf x})$}
\put(1,59){\rotatebox{90}{$\omega_k/\sum \omega_j$}}
\put(16,50){$\Im \{ \omega_k \}$}
\put(35,38){$\Re \{ \omega_k \}$}
\put(36,53){$k$}
\put(16,8){Company 1}
\put(60,8){Company 2}
\put(26,3){DMD mode \#}
\put(70,3){DMD mode \#}
\put(78,31){${\bf x}$}
\end{overpic}
\end{center}
\caption{\label{fig2} DMD decomposition of portfolio data in the transport and home construction (18 companies):
'FDX' 'UNP' 'KSU' 'NSC' 'UPS' 'CHRW' 'UAL' 'DAL' 'LUV' 'CSX' 'DHI' 'PHM' 'TOL' 'NVR' 'HD' 'LOW' 'SHW' 'SPY'.
The panels represent the same information as in Fig.~\ref{fig1} for these two new sectors.
}
\end{figure}

\begin{figure}[t]
\begin{center}
\begin{overpic}[width=0.9\textwidth]{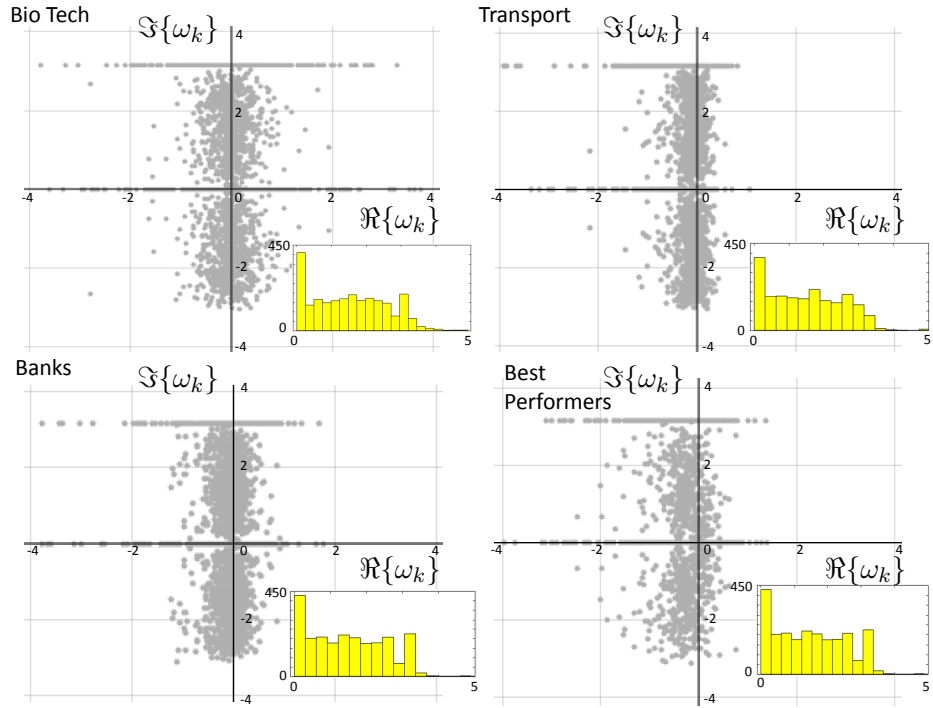}
\normalsize
\put(18,68){$\Im \{ \omega_k \}$}
\put(39,50){$\Re \{ \omega_k \}$}
\put(62,68){$\Im \{ \omega_k \}$}
\put(83,50){$\Re \{ \omega_k \}$}
\put(62,35){$\Im \{ \omega_k \}$}
\put(83,17){$\Re \{ \omega_k \}$}
\put(18,35){$\Im \{ \omega_k \}$}
\put(39,17){$\Re \{ \omega_k \}$}
\end{overpic}
\end{center}
\caption{\label{fig3} DMD distribution of eigenvalues ($\omega_k$) over a 10 year period in 
three key market segments:  biotech ('BIIB' 'AMGN' 'CELG' 'GILD' 'REGN' 'VRTX' 'ALXN' 'ILMN' 'SPYÕ), transportation and home construction ('FDX' 'UNP' 'KSU' 'NSC' 'UPS' 'CHRW' 'UAL' 'DAL' 'LUV' 'CSX' 'DHI' 'LEN' 'PHM' 'TOL' 'NVR' 'HD' 'LOW' 'SHW' 'SPYÕ ), and banks ('JPM' 'MS' 'DB' 'CS' 'BAC' 'CÕ 'BCS' 'BX' 'GS' 'SPYÕ).  Additionally, we
have included the best performers in our data set from this 10 year period ('PCYC' 'REGN' 'GMCR' 'MDVN' 'PCLN' 'NFLX' 'SPYÕ).  Each market segment has a different distribution of eigenvalues that are clustered along the imaginary
axis.  Of importance is the distribution of eigenvalues with $\Re\{\omega_i\}>0$, i.e. those segments that represent the largest growth possibilities and/or volatility.   The  insets of each figure represent a histogram of the distribution
of the magnitude of the eigenvalues for each segment.  The histograms highlight that you cant tell the directionality of markets by looking at the eigenvalue magnitude, however sector volatility may be possible.  Additionally, one
can characterize, via the eigenvalues, differences between rapidly changing sectors versus slower changing sectors. 
Note that vertical structure in the distribution of eigenvalues that is generated by the finite sampling window chosen.}
\end{figure}

\begin{figure}[t]
\begin{center}
\begin{overpic}[width=0.95\textwidth]{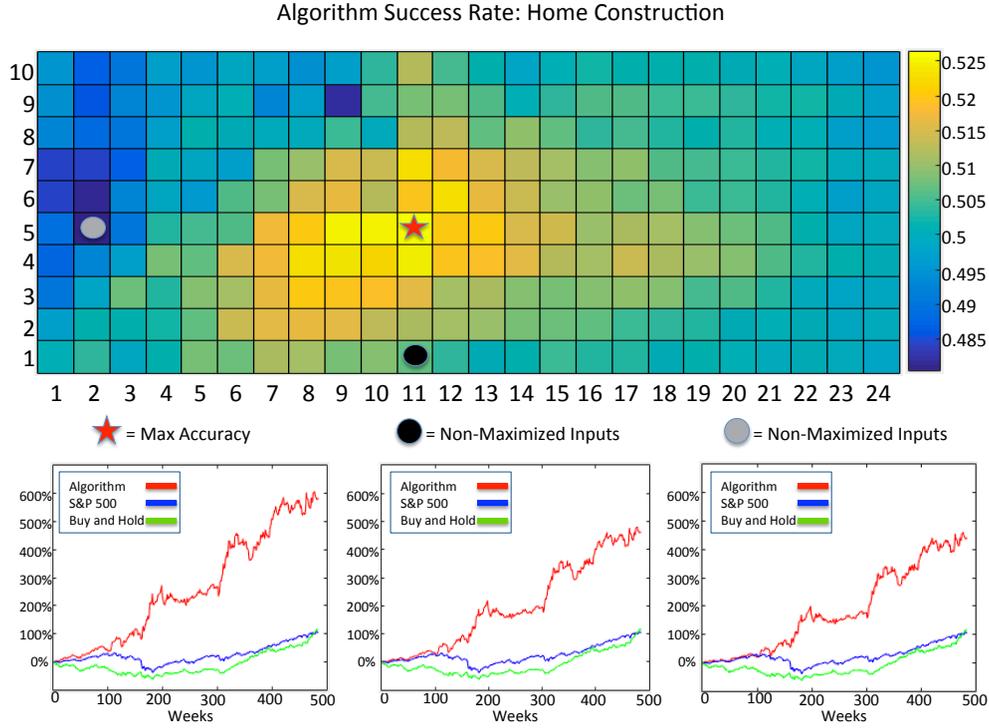}
\normalsize
\end{overpic}
\end{center}
\vspace*{-0.3in}
\caption{\label{fig4} Success rates achieved using different sampling and prediction
windows $(m,\ell)$ for the DMD algorithm when back testing.  The hot-spot (top panel) shows a success rate of about 52.5\% over the last 10 years, meaning that 52.5\% of all the trades we executed were correct and made money.   
The bottom panels show how much money the algorithm makes if we use different inputs for the DMD in comparison to the S\&P 500 as well as a buy-and-hold on the stocks used in the DMD algorithm. Using the best
hot-spot  ${\bf x}_{\mbox{\tiny DMD}} (11,5)$ gives the best return of ~21.48\% annualized over 10 years, however using other inputs such as ${\bf x}_{\mbox{\tiny DMD}} (11,1)$ or ${\bf x}_{\mbox{\tiny DMD}} (2,5)$ we still get promising results of ~19.22\% and ~18.59\% annualized respectively.  When calculating the money we made we ran the DMD and traded off its signal each day entering a position, either long or short, on every company in the algorithm. In the case above we used 10 companies that are in the home construction sector ('DHI' 'LEN' 'PHM' 'TOL' 'NVR' 'HD' 'LOW' 'SHW' 'SPY'). }
\end{figure}

\begin{figure}[t]
\begin{center}
\begin{overpic}[width=0.95\textwidth]{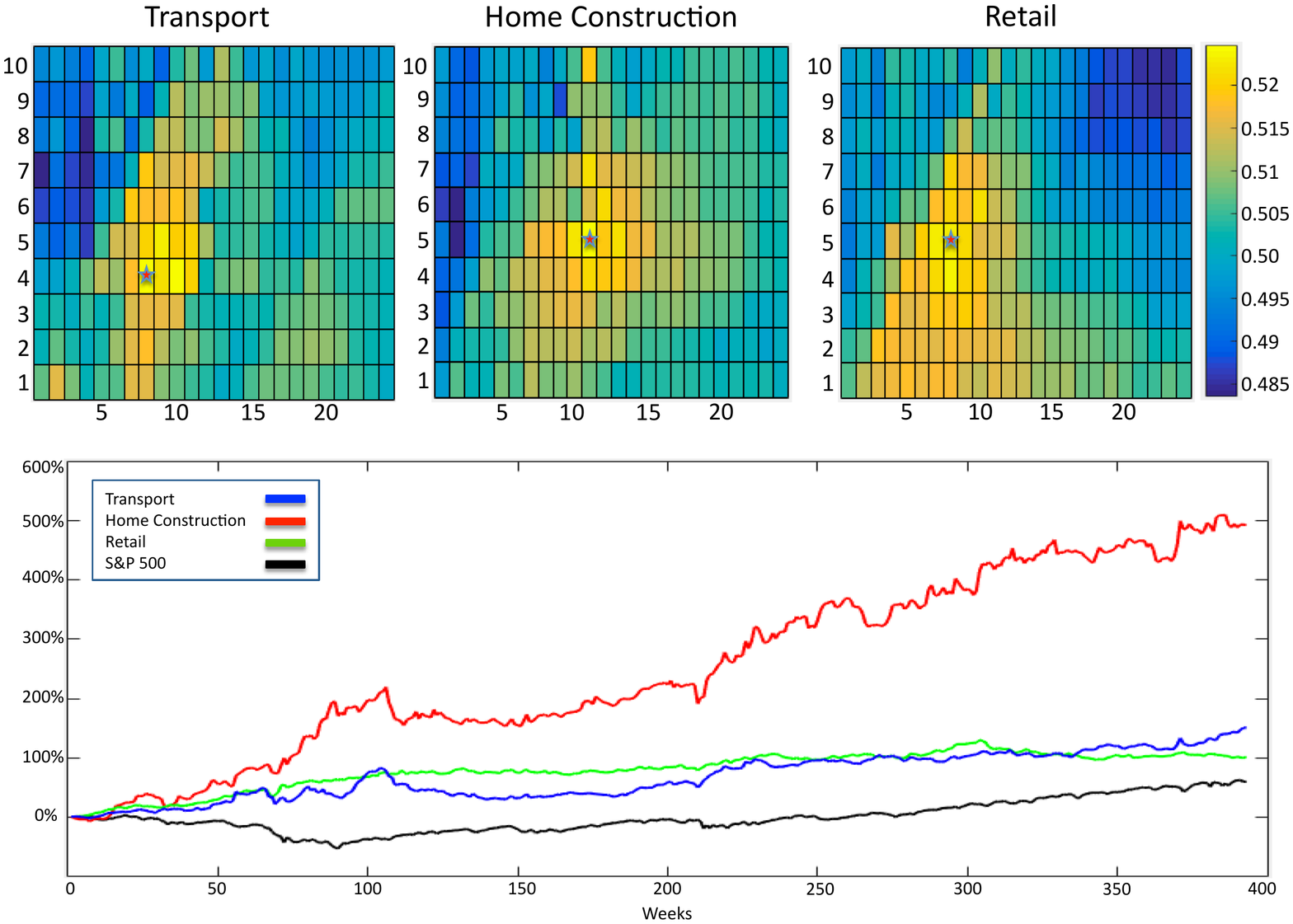}
\normalsize
\end{overpic}
\end{center}
\vspace*{-0.3in}
\caption{\label{fig5} This figure highlights how the algorithm does across sectors with hot-spots vs the benchmark index, the S\&P 500. The adaptive, machine learning, aspect of the trading alg allows for different hot-spots and it chooses the hot-spots that best fit the historic data.  All three sectors in this example have obvious hotspots with high success rates
using different sampling and predictive windows, however they all outperform the benchmark over the last 8 years.  It is especially promising to see the lack of market correlation during the financial crisis, when the benchmark decreased but all trading algs made money.  The three sectors considered are (i) transportation ${\bf x}_{\mbox{\tiny DMD}} (8,4)$:  'FDX' 'UNP' 'KSU' 'NSC' 'UPS' 'CHRW' 'UAL' 'DAL' 'LUV' 'CSX' 'SPY', (ii) home construction ${\bf x}_{\mbox{\tiny DMD}} (11,5)$:  'DHI' 'LEN' 'PHM' 'TOL' 'NVR' 'HD' 'LOW' 'SHW' 'SPY' , (iii) retail ${\bf x}_{\mbox{\tiny DMD}} (8,5)$:
'NFLX' 'KR' 'AMZN' 'CVS' 'WBA' 'TGT' 'COST' 'PCLN' 'SPYÕ.}
\end{figure}

\subsection{DMD Decomposition in Sectors}

To illustrate the DMD decomposition, we consider first a sample of 17 companies over an 18-day trading
window from the
biotech and healthcare sectors:  'DHI' 'LEN' 'PHM' 'TOL' 'NVR' 'HD' 'LOW' 'SHW' 'ONCS' 'BIIB' 'AMGN' 'CELG' 'GILD' 'REGN' 'VRTX' 'ALXN' 'ILMN'.  Figure~\ref{fig1} shows all aspects of the resulting decomposition.  Specifically,
the singular values decomposition (\ref{eq:kry3}) produces a diagonal matrix whose entries determine the
modes of maximal variance.  In this case, there a low-rank structure to the data as demonstrated in Fig.~\ref{fig1}.
The first mode is particularly important as is represents the average price across the sampling window.
The first four DMD modes, which are composed of weightings of the 17 companies, are highlighted as they are the most dominant structures in the data.  The
distribution of eigenvalues $\omega_k$ are also illustrated, showing the modes which have growth, decay
and/or oscillatory behavior.  The DMD scheme takes advantage of identifying the largest growth modes for
investment purposes.  Finally, the weighting of each company unto the DMD modes is also illustrated.
This is the information extracted at each pass of the DMD algorithm for a given data sampling window.
Figure~\ref{fig2} shows the same decomposition in the transport and home construction sectors (18 companies):
'FDX' 'UNP' 'KSU' 'NSC' 'UPS' 'CHRW' 'UAL' 'DAL' 'LUV' 'CSX' 'DHI' 'PHM' 'TOL' 'NVR' 'HD' 'LOW' 'SHW' 'SPY'.
Importantly, one can see the difference in the growth modes between the sectors.  This will
be important for predicting successful sectors.

Given the importance of the eigenvalues, we consider the
DMD distribution of eigenvalues ($\omega_k$) over a 10 year period in 
three key market segments:  biotech ('BIIB' 'AMGN' 'CELG' 'GILD' 'REGN' 'VRTX' 'ALXN' 'ILMN' 'SPYÕ), transportation and home construction ('FDX' 'UNP' 'KSU' 'NSC' 'UPS' 'CHRW' 'UAL' 'DAL' 'LUV' 'CSX' 'DHI' 'LEN' 'PHM' 'TOL' 'NVR' 'HD' 'LOW' 'SHW' 'SPYÕ ), and banks ('JPM' 'MS' 'DB' 'CS' 'BAC' 'CÕ 'BCS' 'BX' 'GS' 'SPYÕ).  
Figure~\ref{fig3} shows these eigenvalues, showing the clustering of eigenvalues along the imaginary 
axis.  The most important modes are those with the largest real part of the eigenvalue.  
For each sector considered, the magnitude of the real part is plotted in a histogram, showing the potential for
growth in each sector.

\subsection{Performance Evaluation}

The DMD decomposition can be used for trading by taking advantage of identifiable hot-spots in 
${\bf x}_{\mbox{\tiny DMD}} (m,\ell)$.  As already outlined in the learning algorithms section, 
the optimal $m$ and $\ell$ can be identified and investments made accordingly.  As a specific
example, Fig.~\ref{fig4}
shows the hot-spot generated in the home construction sector.  
The hot-spot has a success rate of about 52.5\% over the last 10 years, meaning that 52.5\% of all the trades we executed were correct and made money.   The bottom panels in this figure 
shows how much money the algorithm makes if we use different inputs for the DMD. Using the best
hot-spot  ${\bf x}_{\mbox{\tiny DMD}} (11,5)$ gives the best return of ~21.48\% annualized  over 10 years, however using other inputs such as ${\bf x}_{\mbox{\tiny DMD}} (11,1)$ or ${\bf x}_{\mbox{\tiny DMD}} (2,5)$ we still get promising results of ~19.22\% and ~18.59\% annualized respectively.  When calculating the money we made we ran the DMD and traded off its signal each day entering a position, either long or short, on every company in the algorithm. In the case above we used 10 companies that are in the home construction sector ('DHI' 'LEN' 'PHM' 'TOL' 'NVR' 'HD' 'LOW' 'SHW' 'SPY').

\begin{figure}[t]
\begin{center}
\begin{overpic}[width=0.95\textwidth]{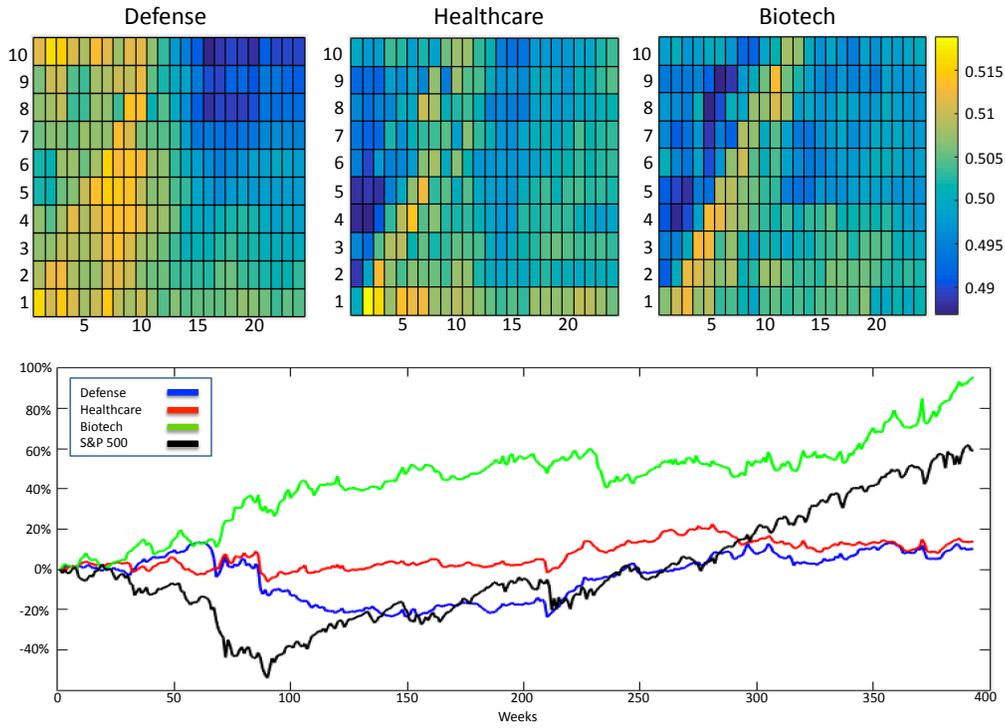}
\normalsize
\end{overpic}
\end{center}
\vspace*{-0.3in}
\caption{\label{fig6} This figure highlights how the algorithm does across sectors lacking trading hot-spots vs the benchmark index, the S\&P 500.  Without a clear hot-spot, it is difficult for the DMD algorithm to find a consistent sampling window for successful predictions.  Indeed, the trading alg does not perform nearly as well as when
a clear hot-spot exists.  There is still a low correlation between the alg and the benchmark during times of crisis, and the alg has a lower volatility as well.   The three sectors considered are (i) defense: 'BA' 'LMT' 'UTX' 'HON' 'GD' 'NOC' 'RTN' 'COL' 'SPY', (ii) healthcare   'JNJ' 'PFE' 'MRK' 'GILD' 'AMGN' 'UNH' 'MDT' 'BMY' 'SPY', (iii)  biotech: 
'BIIB' 'AMGN' 'CELG' 'GILD' 'REGN' 'VRTX' 'ALXN' 'ILMN' 'SPYÕ.
}
\end{figure}

We can expand the performance evaluation to a number of other sectors that exhibit clear and
identifiable trading hot-spots.   Figure~\ref{fig5} shows the application of the  method to the
transport, home construction, and retail sectors.  These three sectors show a clear and pronounced
hot-spot at ${\bf x}_{\mbox{\tiny DMD}} (8,4)$, ${\bf x}_{\mbox{\tiny DMD}} (11,5)$ and
${\bf x}_{\mbox{\tiny DMD}} (8,5)$ respectively.  If our trading strategy is applied in these sectors,
the returns are indicated in the bottom panel of this figure.  The hot-spots allow one to
beat the S\&P 5000 in  all cases.  In contrast, when no hot-spots are evident, as demonstrated
in Fig.~\ref{fig6}  for the defense, healthcare and biotech sectors, there is no longer a guarantee
about performance as the algorithm no longer has a clear trading target (hot-spot) to track.
Note that in our third trading alg proposed, a lack of a clear hot-spot would result in a holding
strategy, thus avoiding trading with no clear signals.

\begin{figure}[t]
\begin{center}
\begin{overpic}[width=0.95\textwidth]{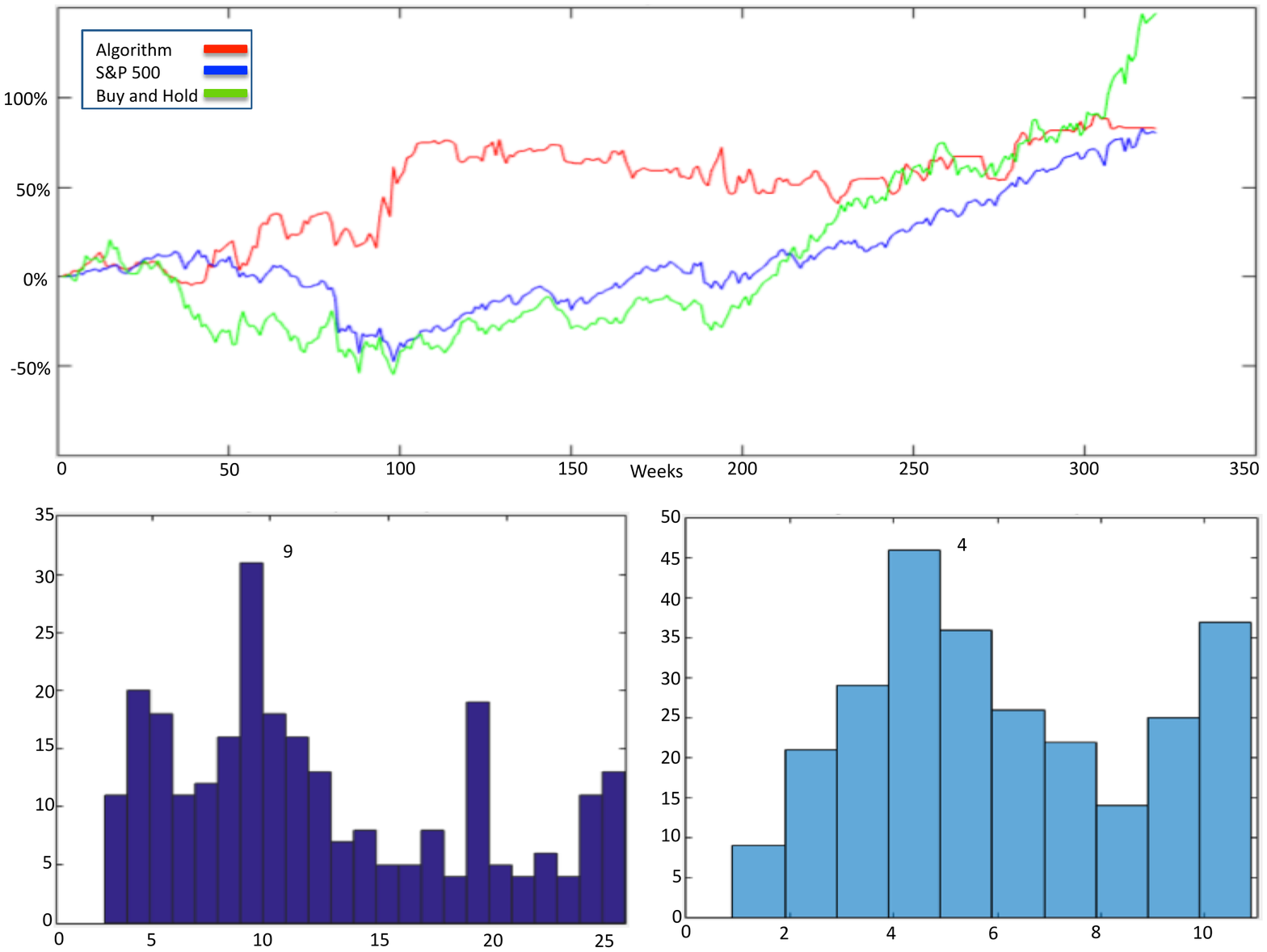}
\normalsize
\end{overpic}
\end{center}
\vspace*{-0.3in}
\caption{\label{fig7} This figure shows how the algorithm does when we change the sampling window and prediction window as time evolves. In this example the alg looks at the last 100 days and finds the inputs with the highest success rate. It then looks to see if it is a hotspot, by evaluating if the 3x3 square, with highest success in the center, has a mean of greater than 53\%. If there is a hot-spot, it will trade.  Otherwise it holds the position. We chose to do this because, as shown in previous figures, the alg works significantly better when hot-spots are detected. Under this algorithm, we traded 83\% of the time. The bar graphs on the bottom shows a histogram of the sampling input window used for trading, and the peaks around a 9-day sampling window.  Also shown is a histogram of the
prediction window with a peak 4-days.  The ${\bf x}_{\mbox{\tiny DMD}} (9,4)$ is consistent with the findings
of Fig.~\ref{fig5}.
}
\end{figure}

Finally, we consider the adaptation of the trading window as a function of time.  In particular,
the sampling and prediction window change over time and it is crucial for the algorithm to 
adjust ${\bf x}_{\mbox{\tiny DMD}} (m,\ell)$ in order to continually optimize performance according to the
market data.  Figure~\ref{fig7} demonstrates the sampling histogram for both the sampling
window and the prediction window.  The learning algorithm changes the sampling window and prediction window by using the last 100 days of data  and finding the ${\bf x}_{\mbox{\tiny DMD}} (m,\ell)$ with the highest success rate. It then looks to see if it is a hot-spot, by evaluating if the 3x3 square, with highest success in the center, has a mean of greater than 53\%. If there is a hot-spot, it will trade.  Otherwise it holds the position. We chose to do this because, as shown in previous figures, the alg works significantly better when hot-spots are detected. Under this algorithm, we traded 83\% of the time. The histogram of the sampling input window used for trading peaks around a 9-day sampling window whereas the
prediction window peaks at 5-days.  The histogram generated ${\bf x}_{\mbox{\tiny DMD}} (9,5)$ is consistent with the findings of Fig.~\ref{fig5}.  This adaptability is an attractive feature of the method as the method
can self-tune in order to improve its efficacy.

\section{Conclusions and Outlook}
\label{sec:conclusion}

Data-driven strategies for analyzing complex systems such as market behaviors and fluctuations 
are of growing interest in the mathematical sciences.   Indeed, the integration of traditional
data modeling methods from statistics and dynamical systems theory can provide enabling strategies
for exploiting patterns of activity in market data.  The dynamic mode decomposition presented
here, which has been successfully implemented in many areas of the engineering, physical and 
biological sciences, provides the foundation of a robust and adaptive trading strategy
that capitalizes on patterns in market activities. 

Overall, The DMD method provides a decomposition of data into
a set of dynamic modes that are derived from snap shots or measurements of a portfolio over a
given time period. 
The DMD method approximates the modes of the so-called Koopman operator. The Koopman operator is a linear, infinite-dimensional operator that represents nonlinear, infinite-dimensional dynamics without linearization, and is the adjoint of the Perron-Frobenius operator. The method can be viewed as computing, from the experimental data, the eigenvalues and eigenvectors (low-dimensional modes) of a linear model that approximates the underlying dynamics, even if the dynamics is nonlinear.  By interpreting the DMD eigenvalues as corresponding to prescribed time scale dynamics, one can
extract coherent data structures in the data.

One can envision a number of innovations to augment the proposed DMD strategy for finance.  Many are
particularly attractive for applications across the engineering, physical and biological sciences, and there is
no reason to believe they would not also be effective in financial settings.  
Such techniques make use of compressive sampling~\cite{cdmd} of market data to facilitate the collection of considerably fewer measurements. This reduction in the number of measurements may have a broad impact in situations where data acquisition is expensive and/or prohibitive.  A second important direction revolves around recent innovations
of the DMD with control~\cite{dmdc}, which is capable of disambiguating between the underlying dynamics and the effects of actuation, or external market drivers, resulting in accurate input-output models. The method is data-driven in that it does not require knowledge of the underlying governing equations, only snapshots of state and actuation data from historical, experimental, or black-box simulations.  One can envision developing such input-output models in various market sectors.  Finally,  modern tools of statistical analysis and dimensionality-reduction have become the 
workhorses for the burgeoning field of machine learning (ML).   ML techniques aim to capitalize on
underlying low-dimensional patterns and clustering in data.  In the dynamical applications considered here, one might exploit these patterns, or DMD modes, by building libraries of low-rank dynamical modes, much like is done with POD modes~\cite{bright2013,brunton2014,epj2014}.  Such DMD libraries for different dynamical regimes partner nicely with compressive sensing strategies.  Additionally, Kernel based techniques, which are at the core
of support vector machines, for instance, have already found
successful application in the DMD architecture when considering more accurate, nonlinear dynamical 
reconstructions~\cite{williams2015}.  Maximum advantage should be
taken of such techniques when integrating the DMD architecture in applications.

\section*{Acknowledgment}
We are especially grateful for discussions with Steven Brunton, Joshua L. Proctor and Jonathan Tu.  

\bibliographystyle{unsrt}
\bibliography{mrdmd.bib}
\end{document}